\documentclass[
 reprint,
longbibliography,
 aps,
 twoside,
 superscriptaddress,
bibnotes
]{revtex4-2}
\usepackage{amsmath, amssymb, graphicx, bm}
\usepackage{bbm}
\usepackage{xcolor}
\usepackage{appendix}
\usepackage{braket}
\usepackage{multirow}
\usepackage{tikz}
\usepackage[hidelinks]{hyperref}
\usepackage{etoolbox}

\newcommand{\ketbra}[2]{\ket{#1} \! \bra{#2}}

\newcommand{\Proj}{\hat{\Pi}}
\renewcommand{\d}{\vec{d}}
\newcommand{\clicks}{C}
\newcommand{\vacs}{V}
\newcommand{\Uhat}{\hat{\mathcal{U}}}
\newcommand{\fock}{\vec{n}}
\newcommand{\fockout}{\vec{m}}
\newcommand{\ii}{\mathrm{i}}
\newcommand{\bigO}{\mathcal{O}}
\newcommand{\al}{\vec{\alpha}}
\newcommand{\gl}{\vec{\gamma}}
\newcommand{\pr}[1]{| #1 \rangle \langle #1|}
\newcommand{\cov}{\Sigma}
\newcommand{\om}{O}
\DeclareMathOperator{\tr}{tr}
\DeclareMathOperator{\per}{per}

\DeclareMathOperator{\lhaf}{lhaf}
\DeclareMathOperator{\tor}{tor}
\DeclareMathOperator{\ltor}{ltor}
\DeclareMathOperator{\brs}{brs}
\DeclareMathOperator{\ubrs}{ubrs}

\begin{document}

\title{Threshold detection statistics of bosonic states}

\author{J. F. F. Bulmer}
\email[email: ]{jakefranklinbulmer@gmail.com}
\altaffiliation[Current affiliation: ]{PsiQuantum, Palo Alto, CA, United States}
\affiliation{Quantum Engineering Technology Laboratories, University of Bristol, Bristol, BS8 1FD, UK}

\author{S. Paesani}
\affiliation{Center for Hybrid Quantum Networks (Hy-Q), Niels Bohr Institute, University of Copenhagen,
Blegdamsvej 17, DK-2100 Copenhagen, Denmark}

\author{R. S. Chadwick}
\affiliation{Quantum Engineering Technology Laboratories, University of Bristol, Bristol, BS8 1FD, UK}
\affiliation{Quantum Engineering Centre for Doctoral Training, University of Bristol, Bristol BS8 1FD, UK}

\author{N. Quesada}
\affiliation{Department of Engineering Physics, \'Ecole Polytechnique de Montr\'eal, Montr\'eal, QC, H3T 1JK, Canada}

\begin{abstract}
In quantum photonics,
threshold detectors,
distinguishing between vacuum and one or more photons,
such as superconducting nanowires and avalanche photodiodes,
are routinely used to measure Fock and Gaussian states of light.
Despite being the standard measurement scheme, 
there is no general closed form expression for measurement probabilities with threshold detectors, 
unless accepting coarse approximations or combinatorially scaling summations.
Here, we present new matrix functions to fill this gap.
We develop the \textit{Bristolian} 
and the \textit{loop Torontonian}
functions for threshold detection
of Fock and displaced Gaussian states, 
respectively,
and connect them to each other and to existing matrix functions. 
By providing a unified picture of bosonic statistics 
for most quantum states of light, we provide novel tools for the design and analysis of photonic quantum technologies. 
\end{abstract}

\maketitle

\section{Introduction}

\begin{figure}[t]
    \centering
    \includegraphics[width=\columnwidth]{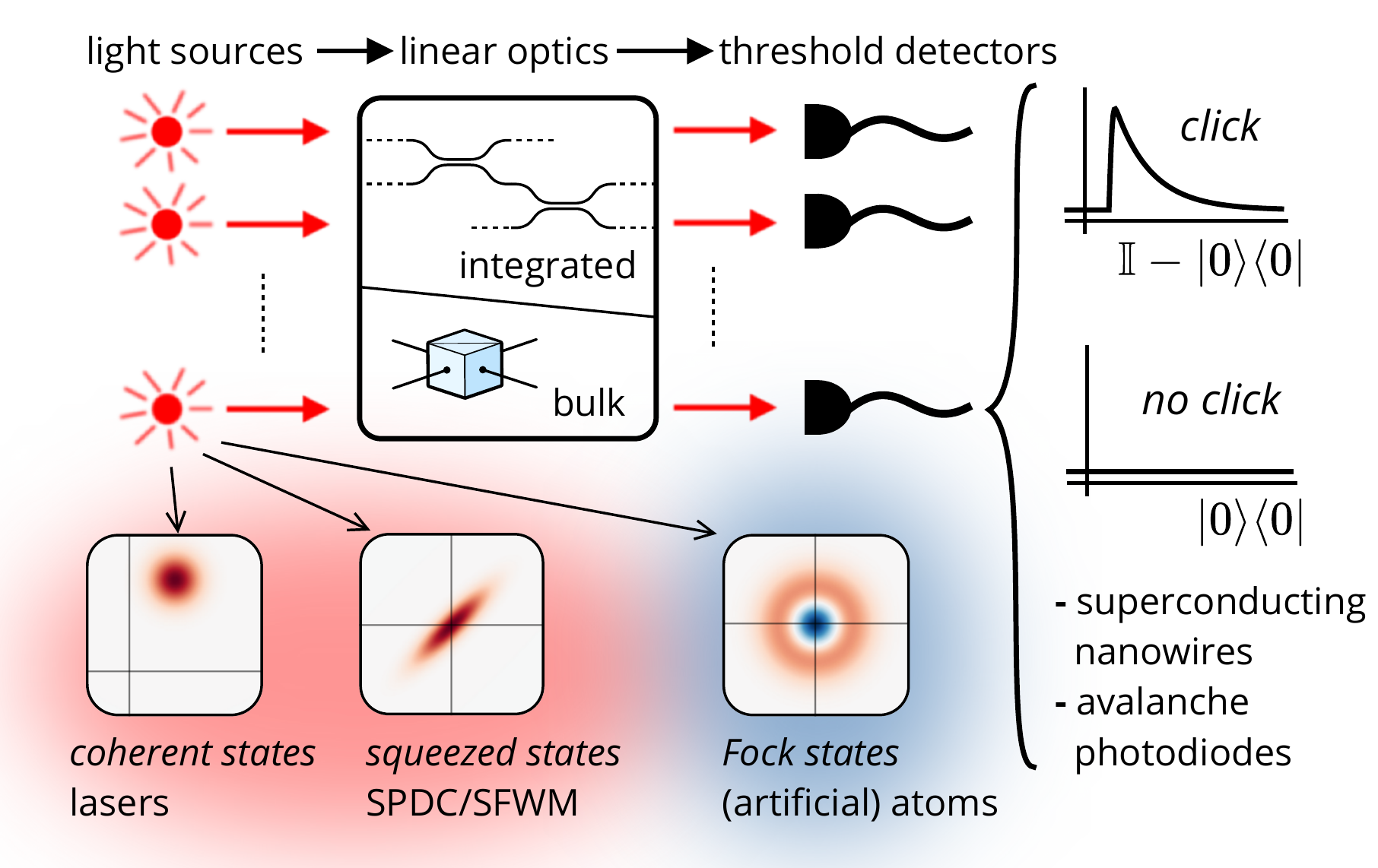}
    \caption{Types of typical quantum photonic experiments which can be modelled using the results of this work. 
    Threshold detection statistics of Gaussian states (highlighted in red) formed by squeezing, displacement and linear optics are captured by the loop Torontonian matrix function. 
    Displacement can be generated using coherent states from lasers, 
    squeezing can come from nonlinear processes such as Spontaneous Parametric Down Conversion (SPDC) or Spontaneous Four-Wave Mixing (SFWM).
    Linear optics can be implemented in a variety of platforms, including bulk and integrated optics.
    States created by the linear optical interference of Fock states (highlighted in blue), as generated by (artificial) atoms, lead to threshold detection probabilities given by the Bristolian matrix function.}
    \label{concept figure}
\end{figure}

Quantum photonic experiments can generally be described as preparing quantum states of light, evolving them through linear optical interferometers, and detecting the output photons.
While the most common types of photonic states, Fock states and Gaussian states, can be routinely prepared via spontaneous processes in optical non-linearities or (artificial) atomic systems and processed with high-fidelity linear optical components~\cite{flamini2018}, photon number detection is typically approximated via the use of threshold photon detectors.
Threshold detection, i.e. a measurement distinguishing only between vacuum and the presence of one or more photons, is widely available, e.g. via high-efficiency superconducting nanowires~\cite{reddy2020superconducting} or room-temperature avalanche photodiodes~\cite{warburton2009free}, making it the standard measurement apparatus in quantum photonics.  
Its use in experiments, represented in Fig.~\ref{concept figure}, encompasses many areas of quantum research, including demonstrations of quantum advantages, e.g. in computation~\cite{zhong2021phase}, measurement sensitivity~\cite{slussarenko2017unconditional}, and loophole-free tests of non-locality~\cite{shalm2015strong}.
However, threshold detection only provides a meaningful approximation of the desired Fock basis measurement projectors in the regime of low mean photon numbers per mode.  
On the other hand, as technology progresses, mean photon numbers increase and higher fidelities are demanded~\cite{zhong2021phase}, making this approximation less appropriate.

To circumvent this issue, experiments can be described directly using the output statistics of threshold detection instead of its photon number resolving approximation. 
However, despite the wide adoption of such systems, there are in general no closed form expressions in the literature for computing measurement probabilities of threshold detectors.
In fact, while an expression for the threshold detection of zero-displaced Gaussian states is known, given by the Torontonian matrix function~\cite{quesada2018gaussian}, no analogous expressions exist for other commonly used states, e.g. Fock or displaced Gaussian states.
For example, for Fock states with fixed photon number, threshold probabilities could be exactly calculated by summing over all possible output states which lead to the given threshold detector outcome. 
However, this method requires calculating a number of probabilities scaling combinatorially with the number of clicked detectors, rendering it impractical already for smaller-scale experiments~\cite{wang2018toward, wang2019boson, thekkadath2022experimental}.
New methods are required to describe quantum photonic technologies which use threshold detection. 
Here, we provide such methods by developing a unified picture to compute threshold statistics for most quantum photonic states of experimental interest.
As described in Table~\ref{table}, this is achieved  by introducing two new matrix functions, the \textit{Bristolian} and the \textit{loop Torontonian}, for threshold statistics with Fock and displaced Gaussian states, respectively, and demonstrating close connections between them and to other existing matrix functions.
The developed tools provide exact simulation, design, and analysis methods for current~\cite{bentivegna2015, wang2018toward, wang2019boson, paesani2019, zhong2021phase, thekkadath2022experimental} and
future quantum photonic systems that use threshold detection.
We wish to highlight the different challenges between computing probabilities, known as \textit{strong simulation}, which we focus on in this work, and drawing samples from a probability distribution, known as \textit{weak simulation}~\cite{van2010classical}. 
These tasks often have very different complexity.
For example, using methods from Ref.~\cite{bulmer2021boundary}, we can sample threshold detector outcomes without ever calculating a threshold detection probability.
\begin{table}[t]
\begin{tabular}{c|cc}
\hline \hline 
\multirow{2}{*}{\textbf{State}} & \multicolumn{2}{c}{\textbf{Detector}} \\ 
                         & number resolving & threshold         \\ \hline
Fock               & permanent       & Bristolian*       \\ 
zero-mean Gaussian & Hafnian         & Torontonian       \\
displaced Gaussian & loop Hafnian    & loop Torontonian* \\ \hline \hline
\end{tabular}
\caption{Matrix functions for the calculation of measurement probabilities in quantum photonics. (*)-symbol is used to indicate functions which are introduced in this work.}
\label{table}
\end{table}

\section{Threshold detection statistics from vacuum statistics}
Threshold detectors are described by the measurement operators
\begin{subequations}
\begin{align}
\Proj_j^{(0)} &= \ketbra{0_j}{0_j}, \\
\Proj_j^{(1)} &= \sum_{k=1}^\infty \ketbra{k_j}{k_j} = \mathbb{I} - \ketbra{0_j}{0_j} \label{click_op},
\end{align}
\end{subequations}
for vacuum (0) and click (1) outcomes on a mode described by label $j$. We use $\ket{0}$ to denote the vacuum state of an optical mode, $\ket{k} = (\hat{a}^\dagger)^k \ket{0}/ \sqrt{k!}$ for Fock states of the optical mode, and $\mathbb{I}$ is the identity operator (we will always assume its dimension to be the same as the other operators appearing in the equation).

We write the outcome of $M$ threshold detectors, labelled with $ j\in[M] = \{1,2,\dots,M\}$, using a length-$M$ bit-string $\d$, where the $j$th element gives the measurement outcome of the $j$th mode. 
Defining a set of modes which clicked, $\clicks = \{j\in [M] \ |\ d_j=1\}$, and a set for modes with the vacuum outcome, $\vacs = \{j\in [M]\  |\ d_j =0\}$, we can write the multimode measurement operator as 
\begin{align}
    \Proj^{(\d)} & =  \bigotimes_{j=1}^M \Proj^{(d_j)}_j = \bigotimes_{j \in \clicks} \left(\mathbb{I} - \ketbra{0_j}{0_j}\right) 
    \bigotimes_{k \in \vacs} \ketbra{0_k}{0_k},
\end{align}
which can be rearranged to give 
\begin{align}
    \Proj^{(\d)} = \ketbra{\vec{0}_\vacs}{\vec{0}_\vacs}
    \sum_{Z \in P(\clicks)} (-1)^{|Z|} \ketbra{\vec{0}_{Z}}{\vec{0}_{Z}}.
    \label{incexc}
\end{align}
Here, we use $P(\clicks)$ to denote the powerset of $C$ and $|Z|$ for the number of elements in a set $Z$.
$\ketbra{\vec{0}_\vacs}{\vec{0}_\vacs}$ describes the vacuum projector in all the vacuum outcome modes and $\ketbra{\vec{0}_{Z}}{\vec{0}_{Z}}$ describes the vacuum projector in all modes in a subset $Z\subseteq \clicks$.

Eq.~\eqref{incexc} indicates that to calculate the threshold detection probabilities for any state it is sufficient to calculate marginal vacuum probabilities, which are used in an inclusion/exclusion sum as described by Eq.~\eqref{incexc}. 
Using this measurement operator and the Born rule $p(\d) = \tr(\Proj^{(\d)} \rho)$ on some state $\rho$, we find:

\begin{align}
    p(\vec{d}) = 
    \sum_{Z \in P(C)} (-1)^{|Z|} p(\d_\vacs=\vec{0}, \d_Z=\vec{0})
    \label{thresh_vac},
\end{align}
where $p(\d_V=\vec{0}, \d_Z=\vec{0})= \tr(\ketbra{\vec{0}_\vacs}{\vec{0}_\vacs}\otimes\ketbra{\vec{0}_{Z}}{\vec{0}_{Z}} \rho)$.
This formula provides our starting point for deriving general expressions for threshold detection statistics.

\section{Marginal vacuum probabilities from the photon number probability generating function}
For an $M$-mode linear optical interferometer, described by an $M \times M$ matrix $U$ and the operator $\Uhat$,  the creation operators are transformed as 
\begin{align}
  \Uhat \hat{a}^\dagger_j \Uhat^\dagger = \sum_{k=1}^M U_{kj} \hat{a}^\dagger_k.
\end{align}
Considering an input state $\ket{\Phi_0}$, the output photon number probability distribution is then
\begin{align}
    p(\fockout) = \left| \bra{\fockout}\Uhat\ket{\Phi_0} \right|^2
    \label{fock_probs}
\end{align}
where $\fockout$ is a length-$M$ list describing the photon number in each mode at the output of the interferometer, and 
$\ket{\fockout} = \bigotimes_{j=1}^M \left((\hat{a}^\dagger_j)^{m_j}/\sqrt{m_j!}\right)\ket{0}$.

Following Ref.~\cite{ivanov2020complexity}, considering the Fourier transform of the probability distribution of photon number basis measurements we define the characteristic function 
\begin{align}
    \chi(\vec{\phi}) = \sum_{\fockout} \exp \left(\ii \sum_{j=1}^M \phi_j m_j \right) p(\fockout)
    \label{char_def}
\end{align}
which, with some manipulation (see Appendix~\ref{deriv}), can be expressed as 
\begin{align}
    \chi(\vec{\phi}) = \bra{\Phi_0} \Uhat^\dagger \ \Uhat_{\vec{\phi}}\  \Uhat \ket{\Phi_0} 
    \label{char}
\end{align}
where $\Uhat_{\vec{\phi}}$ is the operator given by the evolution due to the linear optical transformation 
\begin{equation}
    U_{\vec{\phi}}=\bigoplus_{j=1}^M \exp (\ii \phi_j).
    \label{eq:uphi}
\end{equation}

We can transform this into a probability generating function, $G$, using the substitution $x_j = \exp(\ii \phi_j)$:
\begin{align}
    G(\vec{x}) = \sum_{\fockout} \left( \prod_{j=1}^M x_j^{m_j} \right) p(\fockout).
    \label{prob_gen}
\end{align}

The function $G(\vec{x})$ has the following useful properties.
To marginalise the $j$th mode, we simply set $x_j=1$.
If we set $x_j=0$, this gives us the probability for $n_j=0$. 
Therefore, if we want to calculate the probability that some subset of the modes, $\vacs$, measure vacuum and we marginalise over all other modes, $B$, we can evaluate 
\begin{align}
    p(\fockout_\vacs = \vec{0}) = G(\vec{x}_\vacs = \vec{0}, \vec{x}_B = \vec{1}),
    \label{vac_gen}
\end{align}
where $\vec{0}$ ($\vec{1}$) is a vector with 0 (1) in all entries.
$G(\vec{x})$ is the probability distribution generating function, so by taking derivatives of $G(\vec{x})$, we can find information about the photon number basis probability distribution~\cite{ivanov2020complexity}.

By using the expression for the characteristic function in Eq.~\eqref{char}, we can see that this amounts to calculating the scattering amplitude of $\ket{\Phi_0}$ to itself, through a linear optical interferometer described by the transformation $U^\dagger U_{\vec{\phi}} U$. 
From again using the substitution $x_j = \exp(\ii \phi_j)$ in Eq.~\eqref{eq:uphi}, we see that $U_{\vec{\phi}}$ physically corresponds to either zero transmission for modes in $\vacs$, or unit transmission for modes in $B$.
As we show in the next sections, we can use this, in conjunction with Eq.~\eqref{thresh_vac} to calculate threshold detection probabilities for all the experimental scenarios outlined in Fig.~\ref{concept figure}.

\section{Fock state inputs}
Recall that the scattering amplitudes of Fock states evolved through a lossless interferometer are given by the permanent matrix function~\cite{scheel2004permanents} 
\begin{align}
    \bra{\fockout} \Uhat \ket{\fock} = \frac{\per(U_{\fockout, \fock})}{\sqrt{\prod_{j=1}^M n_j! m_j!}}
    \label{per}
\end{align}
where $U_{\fockout, \fock}$ is constructed from $U$ by repeating its $j$th row $m_j$ times and its $j$th column $n_j$ times for all $j \in [M]$.

Therefore if we have an $N$-photon input Fock state, $\ket{\Phi_0}=\ket{\fock}$, we can use Eq.~\eqref{char}, Eq.~\eqref{prob_gen} and Eq.~\eqref{per} to write 
\begin{align}
    G(\vec{x}) = \frac{\per\left([U^\dagger U_{\vec{x}} U]_{\fock,\fock} \right)}{\prod_{j=1}^M n_j !}
    \label{fock_gen}
\end{align}
where $U_{\vec{x}}$ is formed like $U_{\vec{\phi}}$, but with diagonal matrix elements: $[U_{\vec{x}}]_{jj} = x_j$. 
For this equation to be valid, we must have a lossless unitary transformation.
However, we are free to marginalise over modes by allowing elements of $\vec{x}$ to be set to 1 for any mode we wish to marginalise over, including any loss modes, as shown in Eq.~\eqref{vac_gen}. 

Because Eq.~\eqref{fock_gen} provides us with a closed form expression for marginal vacuum probabilities, and because
Eq.~\eqref{thresh_vac} shows us that marginal vacuum probabilities are sufficient to calculate threshold detection probabilities, we can use this to derive a matrix function for calculating threshold detection probabilities of Fock states.
For more generality, we first consider a linear optical transformation with losses, described by an $M_\text{in} \times M_\text{out}$ matrix $T$, with singular values upper bounded by $1$~\cite{garcia2019simulating}.
In Appendix~\ref{der_brs}
we show that if the input state is an $M_\text{in}$-mode Fock state, $\fock$, then by combining Eq.~\eqref{thresh_vac} with Eq.~\eqref{fock_gen}, 
we can calculate the threshold detection probability of the outcome described by an $M_\text{out}$-length bit-string $\d$ as:
\begin{align}
    p(\d) = \frac{\brs\left(T_{\d,\fock}, E(T)_{\fock,\fock} \right)}{\prod_{j=1}^{M_\text{in
    }} n_j ! }.
\end{align}
Here we have introduced a matrix function, the \textit{Bristolian}, defined as 
\begin{multline}
    \brs\left(A, E \right) = \\
    \sum_{Y \in P([m])} (-1)^{m - |Y|} \per \left(
    [A_Y]^\dagger A_Y + E 
    \right),
    \label{brs_def}
\end{multline}
where $A$ is an $m \times n$ matrix and $E$ is an $n \times n$ matrix.
$A_Y$ denotes selecting the rows of $A$ according to the elements of $Y$, and $[m] = \{1,2,\dots,m\}$.
We have also defined a matrix which accounts for the mixing with vacuum in the environment modes 
\begin{align}
    E(T) &=  \mathbb{I} -T^\dagger T .
\end{align}

Our naming of the Bristolian is inspired by the convention established by the Hafnian and Torontonian matrix functions, which are named after the cities of their discovery.
By noticing that $\mathbb{I} - T^\dagger T$ gives a zero matrix when $T$ is unitary, the $\brs$ function can be simplified when $T$ is a unitary matrix $U$, only requiring the rows of $U$ which correspond to modes with a detector click, providing
\begin{align}
    p(\d) = \frac{\ubrs\left( U_{\d,\fock} \right)}{\prod_{j=1}^M n_j!}.
\end{align}
Here we defined the \textit{Unitary Bristolian} acting on an $m \times m$ matrix, $A$, as 
\begin{align}
    \ubrs(A) = \sum_{Y \in P([m])} (-1)^{m - |Y|}
    \per\left([A_Y]^\dagger A_Y \right).
    \label{ubrs}
\end{align}

\section{Displaced Gaussian state inputs}
Gaussian states are the set of states that have a Gaussian characteristic function. A Gaussian state $\rho$ is uniquely characterized by its vector of means with entries
\begin{align}
\al_i = \text{tr}\left[ \rho \hat{\vec{\zeta}} _i \right],
\end{align}
and its Husimi covariance matrix with entries
\begin{align}
\cov_{i,j} = \tfrac12 \text{tr}\left( \left[\hat{\vec{\zeta}} _i \hat{\vec{\zeta}} _j^\dagger + \hat{\vec{\zeta}} _j^\dagger \hat{\vec{\zeta}} _i \right] \rho \right) - \al_i \al_j^* + \tfrac12 \delta_{i,j},
\end{align}
where we have used a vector of creation and annihilation operators
\begin{align}\label{mode_ordering}
\hat{\vec{\zeta}} = \left(\hat{a}_1,\ldots,\hat{a}_{M}, \hat{a}_1^\dagger,\ldots,\hat{a}_{M}^\dagger \right).
\end{align}

The Husimi function $Q(\vec{r}) = \bra{\vec{r}} \rho \ket{\vec{r}}$ 
maps displacement vectors, $\vec{r}$, to probabilities, so to calculate vacuum probabilities we can evaluate the Husimi function at the origin.
Noting that we can marginalise over modes by deleting all the corresponding elements of $\cov$ and $\al$, we obtain~\cite{Serafini_2017}
\begin{align}
    p(\fockout_\vacs = \vec{0}) &= Q(\vec{r}_\vacs = \vec{0}) \\
    &=\frac{\exp\left(-\frac{1}{2}\al_\vacs^\dagger[\cov_{\vacs \vacs}]^{-1}\al_\vacs \right)}{\sqrt{\det(\cov_{\vacs \vacs})}}.
    \label{eq:gaussvac}
\end{align}
The notation $\cov_{\vacs\vacs}$ and $\al_\vacs$ differs slightly here from the previous section, as now there are two basis vectors for each mode of our system, corresponding to each mode's $\hat{a}$ and $\hat{a}^\dagger$ operator. 
We form $\cov_{\vacs\vacs}$ by selecting both rows/columns of $\cov$ which correspond to each element of $\vacs$ and we form $\al_\vacs$ by selecting both elements of $\al$ corresponding to each element of $\vacs$. 

We can use this to immediately arrive at a threshold detection probability for displaced Gaussian states using Eq.~\eqref{thresh_vac}.
However, here we must invert and compute determinants for square matrices of size $2(|V|+|Z|)$.
It would be preferable if we could reduce these to matrices of size $2|Z|$.
It would also be helpful conceptually to have a formula which can be connected to other relevant matrix functions, the Torontonian~\cite{quesada2018gaussian} and the loop Hafnian~\cite{quesada2019franck}.
Therefore, it is of interest to write this probability in terms of 
\begin{align}
O = \mathbb{I} -  \cov^{-1} \text{ and } \gl = (\cov^{-1}\al)^*.
\end{align}

In Appendix~\ref{ltor_derivation},  we show how Eq.~\eqref{thresh_vac} can be rearranged into the following:
\begin{align}\label{eq:prob_ltor}
    p(\d) = p(\vec{0}) \ltor\left( \om_{CC}, \gl_{C}  \right),
\end{align}
where $C$ is given by the index of the elements of $\d$ where $d_j=1$, so $O_{CC}$ and $\gl_{C}$ are the matrix and vector formed by selecting the rows/columns of $O$ and elements of $\gl$ which correspond to modes which see a detector click.
$p(\vec{0})$ is the probability of detecting vacuum in all modes, and can be calculated using Eq.~\eqref{eq:gaussvac}.
We introduce the \textit{loop Torontonian}, which is defined as 
\begin{multline}
    \ltor\left( \om, \gl \right) = \\
    \sum_{Y \in P([m])} (-1)^{m-|Y|} \frac{\exp\left[ \tfrac12 \gl_Y^t [\mathbb{I} - \om_{YY}]^{-1} \gl_Y^* \right]}{\sqrt{\det(\mathbb{I} - \om_{YY})}},
    \label{eq:ltor_def}
\end{multline}
where $O$ is a $2m \times 2m$ matrix and $\gl$ is a $2m$-length vector.

\section{Connections between matrix functions}
In the limit of no displacement $\al = \gl = \vec{0}$, the exponential terms in the numerator of Eq.~\eqref{eq:ltor_def} becomes 1 and thus $\ltor\left( \om, \vec{0} \right) = \tor\left( \om \right) $, where $\tor$ is the Torontonian function from Ref.~\cite{quesada2018gaussian}. One can show, using the scattershot construction~\cite{lund2014boson}, that the Torontonian and Bristolian are related via the following limit
\begin{multline}
\brs\left(T_{\vec{d},\fock}, E(T)_{\fock,\fock} \right) = \\ \lim_{\varepsilon \to 0 }  (\varepsilon^{-2} - 1)^{N}  \tor\left( O(\varepsilon)_{CC} \right),
\end{multline}
\begin{equation}
O(\varepsilon) = \varepsilon \begin{pmatrix}
	0 & 0 & 0 & T \\
	0 & \varepsilon E(T)^*
	& T^t & 0 \\
	0 & T^* & 0  & 0\\
	T^\dagger &  0 & 0 & \varepsilon E(T)
	\end{pmatrix}.
\end{equation}
where $\fock$ is a bitstring (implying that this identity is only valid for single-photon or vacuum inputs), $N = \sum_{i} \fock_i $ and $C$ is the union of the labels of the modes in which single photons were input into the interferometer and the labels of the modes in which clicks are registered.  
This relation is proven in Appendix~\ref{app:bris_to_tor}. 

As we show in 
Appendix~\ref{lhaf}, the loop Torontonian can also be used as a generating function for the loop Hafnian,
\begin{equation}
    \lhaf(X \om_{\clicks \clicks}, \gl_{\clicks}) = \left.\frac{1}{\ell!} \frac{d^\ell}{d \eta ^{\ell}} \ltor\left( \eta \om_{\clicks \clicks}, \sqrt{\eta} \gl_{\clicks} \right) \right|_{\eta = 0} 
\end{equation}
where $X = \left[\begin{smallmatrix} 0 & \mathbb{I} \\ \mathbb{I} &  0 \end{smallmatrix}\right] $ and $\ell = |C|$.
We use this to derive the trace formula for the loop Hafnian, the fastest known method for computing photon number resolved measurement probabilities on displaced Gaussian states~\cite{bjorklund2019faster, quesada2019franck, quesada2019simulating}.
Because the loop Hafnian of a bipartite graph is given by the matrix permanent~\cite{bjorklund2019faster}, all the matrix functions in Table~\ref{table} can be derived from the loop Torontonian.

We also see a connection between the Bristolian and the permanent when an $N$ photon Fock state results in $N$ threshold detector clicks.
In this case, each threshold detector must have seen exactly 1 photon, so we can describe the measurement operator of each threshold detector click as a single photon projector, which leads to describing the event with permanents, as given by Eq.~\eqref{per}.
In Appendix~\ref{brs_per}, 
we show this link directly by first describing the Unitary Bristolian for $N$ photon, $N$ click events as the permanent of an $N \times N \times N$ 3-tensor~\cite{tichy2015sampling}.

\section{Time complexities}
In Appendix \ref{app:time_comp},  we discuss the time complexities for the Bristolian and the loop Torontonian.
We find that, using the formulae presented in this work, the Bristolian, $\brs(A,E)$, has a time complexity of $\bigO(n2^{n+m})$ for an $m \times n$ matrix $A$ and $n \times n$ matrix $E$ and the loop Torontonian, $\ltor(O, \gl)$,  has time complexity of $\bigO(m^3 2^m)$ for a $2m \times 2m$ matrix $O$ and $2m$-length vector $\gl$.
For the loop Torontonian, this complexity can be reduced using a recursive strategy which exploits Cholesky decomposition~\cite{kaposi2021polynomial}.
We also believe that the Bristolian's time complexity can likely be reduced, and we leave this as an open problem.

\section{Improved accuracy of a threshold detection model} 
To assess the improvements offered by using the correct description of threshold detection over the common approximation of single photon projective measurement, we present two representative examples.
By simulating the probability distribution for 100 different Haar random unitaries in lossy 4 input photon Fock state Boson sampling experiments on mode numbers from 4 to 12, we evaluate the total variation distance (TVD) between probability distributions from the exact model, which uses the Bristolian, and an approximate model, which uses a sum over matrix permanents, as discussed in 
Appendix~\ref{app:brs_using_pers}.
Although the TVD is reduced for higher numbers of modes, the approximation is always 5\% - 12\% removed from the correct distribution.
To test the loop Torontonian, we use experimental data from Ref.~\cite{thekkadath2022experimental}. We see that for the 2 photon distribution for different levels of displacement, the loop Torontonian offers a better match to the experiment of up to 16\%.
See Appendix \ref{app:accuracy} 
for more detail.

\section{Conclusion}
The new methods we have derived, in particular the Bristolian and the loop Torontonian functions, are useful tools to model and analyse a wide variety of quantum photonic experiments and applications.
For example, the Bristolian is relevant to applications including linear-optical quantum computing~\cite{knill2001scheme, kok2007linear, rudolph2017optimistic}, Boson Sampling~\cite{aaronson2011computational, wang2019boson} and quantum communications~\cite{you2021quantum}, commonly based on threshold detection.
The loop Torontonian can be applied to applications including
Gaussian state reconstruction~\cite{thekkadath2022experimental},
measuring graph similarity~\cite{schuld2020measuring},
calculations of vibronic spectra of molecules~\cite{huh2015boson}, and  quantum metrology~\cite{afek2010high}, and has already been applied for evaluating proposed quantum communication protocols~\footnote{The initial inspiration for us to derive the loop Torontonian came from the need to calculate threshold detection statistics for the quantum communication protocols proposed in Ref.~\cite{bacco2021proposal}}. %
To facilitate their use, we provide example calculations of common experimental scenarios in Appendix~\ref{sec:examples}
using the Bristolian and loop Torontonian, and have made available implementations in the open-source Python package \texttt{The Walrus}~\cite{gupt2019walrus}.
Details for the software implementation are provided in Appendix~\ref{app:code}.
The connections that we have shown between the Bristolian and the permanent (Appendix ~\ref{brs_per}),
the loop Torontonian and the loop Hafnian (Appendix~\ref{lhaf}),
and the Bristolian and the Torontonian (Appendix~\ref{app:bris_to_tor})
indicate that these functions can provide a useful mathematical and conceptual tool for a deeper understanding of bosonic statistics in photonic experiments.

\section*{Acknowledgements}
JFFB and RSC acknowledge support from EPSRC (EP/N509711/1, EP/LO15730/1). 
NQ acknowledges support from the Minist\`ere de l'\'Economie et de l'Innovation du Qu\'ebec and the Natural Sciences and Engineering Research Council of Canada. 
SP acknowledges funding from the Cisco University Research Program Fund nr. 2021-234494.
We thank G. S. Thekkadath for useful discussions and sharing experimental data from Ref.~\cite{thekkadath2022experimental}.
NQ thanks S. Duque Mesa, B. Lanthier, D. Leclerc, B. Turcotte, and J. Zhao for valuable discussions. 
We thank G. Morse for implementing the generalisation of the recursive Torontonian formula~\cite{kaposi2021polynomial} to the loop Torontonian, see \href{https://github.com/XanaduAI/thewalrus/pull/332}{pull request (332)} to \texttt{The Walrus}~\cite{gupt2019walrus}.

\appendix

\section{Examples \label{sec:examples}}

We present some examples of how to apply the Unitary Bristolian, Bristolian and loop Torontonian to some representative situations.

\subsection{Lossless Hong-Ou-Mandel}
Our first example is Hong-Ou-Mandel interference~\cite{hong1987measurement} of single photons on a lossless 50/50 beam splitter.
Here, the input state is $\fock=(1,1)$ and
\begin{align}
    U = \frac{1}{\sqrt{2}}\begin{pmatrix}
    1 & 1 \\
    1 & -1
    \end{pmatrix}.
    \label{eq:bs}
\end{align}
The probability of detecting a coincidence is given by 
\begin{align}
    p(\d = (1,1)) = & \ubrs(U).
\end{align}
So, $C = \{1,2\}$.
We expand the $\ubrs$ function as 
\begin{align}
    \ubrs(U) = &
    \per\left(U^\dagger U\right) \\
    & - \per\left([U_1]^\dagger U_1 \right) 
    - \per\left([U_2]^\dagger U_2 \right),
\end{align}
with $U_1 = (1, 1) / \sqrt{2}$ and $U_2 = (1,-1)/\sqrt{2}$ being the 1st and 2nd rows of $U$ respectively.
We also use that $\per\left(U^\dagger U \right) = \per(\mathbb{I}) = 1$, and calculate 
\begin{align}
     \per\left([U_1]^\dagger U_1 \right) &= \per\left[\frac{1}{2}\begin{pmatrix}
     1 & 1 \\
     1 & 1
     \end{pmatrix} \right] &= \frac12, &\\
     \per\left([U_2]^\dagger U_2 \right) &= 
     \per\left[\frac{1}{2}\begin{pmatrix}
     1 & -1 \\
     -1 & 1
     \end{pmatrix} \right] &= \frac12 &,
\end{align}
giving $p(\d = (1,1)) = 0$, as expected.
Here, we have not included the term where $Y$ is the empty set.
In this case, we are considering the permanent of the all zeros matrix, which is zero and so does not contribute.

We can also see that $
    p(\d=(1,0)) = \per\left([U_1]^\dagger U_1 \right) = 1/2$,
and similarly $p(\d=(0,1)) = 1/2$.
These are due to the $(2,0)$ and $(0,2)$ photon number output terms.

\subsection{3-mode zero transmission law \label{ex:ztl1}}
The zero transmission law (ZTL) tells us that many output states of a Fourier transform interferometer are suppressed due to multi-photon interference.

Using $\omega=\exp(-2 i \pi / 3)$, the three mode Fourier transform interferometer is given by 
\begin{align}
    U = \frac{1}{\sqrt{3}}\begin{pmatrix}
    1 & 1 & 1 \\
    1 & \omega & \omega^2 \\
    1 & \omega^2 & \omega
    \end{pmatrix}.
    \label{eq:dft3}
\end{align}
We consider $\fock=(1,1,1)$. 
According to the ZTL, all permutations of the output $\fockout = (2,1,0)$ should be suppressed, and so $\d=(1,1,0)$, due to $\fockout=(2,1,0)$ and $\fockout=(1,2,0)$, is also expected to be suppressed. 
The probability of this threshold detector outcome is given by
\begin{align}
    p(\d=(1,1,0)) =& \ubrs\left(U_{\d,\fock}\right),
\end{align}
where
\begin{align}
    \ubrs\left(U_{\d,\fock}\right) = & 
    \per\left( [U_{(1,2)}]^\dagger U_{(1,2)} \right) \\
    & - \per\left([U_1]^\dagger U_1 \right)
    - \per\left([U_2]^\dagger U_2 \right) \nonumber .
\end{align}
By evaluating these permanents, we find
\begin{align}
    \ubrs\left(U_{\d,\fock}\right) = 
    \frac49 - \frac29 - \frac29  = 0,
\end{align}
showing a suppression as expected.
We can also calculate 
\begin{align}
    p(\d=(1,1,1)) =& \ubrs(U) = \frac13,
\end{align}
which agrees with the prediction that $\ubrs(U) = \left|\per(U)\right|^2$ when $U$ is square, since $\per(U) = -1 / \sqrt{3}$.

\subsection{Lossy Hong-Ou-Mandel}
The Hong-Ou-Mandel effect is preserved under balanced loss. 
In contrast to the previous examples, in this example example, the loss means that we need to use the Bristolian, as the Unitary Bristolian is no longer valid. 
We consider a transmission matrix defined like $U$ in Eq.~\eqref{eq:bs}, but with transmission $\eta$, giving $T = \sqrt{\eta} U$.
In this case 
\begin{align}
    p(\d=(1,1)) =& \brs(T, E)
\end{align}
where $E = (1 - \eta) \mathbb{I}$.
We evaluate the Bristolian to find
\begin{align}
    \brs(T, E) = & \per\left(T^\dagger T + E\right)  \\
    & - \per\left([T_1]^\dagger T_1 + E\right) - 
    \per\left([T_2]^\dagger T_2 + E\right) \nonumber \\
    & + \per(E) \nonumber \\
    = &\ 1 - 2\left( 1 - \eta + \eta^2/2\right)   + (1 - 2\eta + \eta^2)  \nonumber \\
    = &\ 0  \nonumber 
\end{align}
which confirms that the coincidence event is still suppressed under balanced loss.

\subsection{Lossy zero transmission law}
When there is loss, we start to witness threshold detector outcomes which were suppressed in the lossless case.
We repeat the example in section~\ref{ex:ztl1} but adding a transmission $\eta \leq 1$.
So, $T = \sqrt{\eta} U$ for $U$ defined in Eq.~\eqref{eq:dft3}.

Using the Bristolian, we find 
\begin{align}
    p(\d=(1,1,0)) = \brs(T_{\d,\fock}, E(T)_{\fock,\fock}),
\end{align}
where $E(T) = (1 - \eta) \mathbb{I}$, and this gives
\begin{align}
    \brs(T_{\d,\fock}, E(T)_{\fock,\fock}) = & \per\left([T_{(1,2)}]^\dagger T_{(1,2)} + E(T) \right) \\
    & - \per\left([T_{1}]^\dagger T_{1} + E(T)_{\fock,\fock} \right) \nonumber \\
    & - \per\left([T_{2}]^\dagger T_{2} + E(T)_{\fock,\fock} \right) \nonumber \\
    & + \per(E(T)_{\fock,\fock}) \nonumber \\
    = &\ \eta^2 (1 - \eta) / 3 \label{eq:ztl}.
\end{align}
We can check this problem intuitively by considering all the losses to be applied just before the measurement.
In this picture, the outcome $\d = (1,1,0)$ can only occur when we have the state $\fockout=(1,1,1)$ before the losses, then the first two photons are transmitted and the last photon is lost.
Therefore we expect the probability found in Eq.~\eqref{eq:ztl}.
The factor $1/3$ comes from section~\ref{ex:ztl1}, where $p(\fockout=(1,1,1))=1/3$ before any loss is applied.

For our final Bristolian example calculation, we consider a case where we have to repeat columns more than once.
Consider the same $T$ as above, but with the input state $\fock=(1,2,0)$.
We calculate the probability of $\d=(0,1,1)$ (so $C = \{2,3\}$), denoting the elements of $T$ as $t_{jk}$.
\begin{align}
    p(\d=(0,1,1)) = \frac12 \brs\left(T_{\d,\fock}, E(T)_{\fock,\fock} \right),
\end{align}
where,
\begin{align}
    T_{\d,\fock} = & \begin{pmatrix}
        t_{22} & t_{23} & t_{23} \\
        t_{32} & t_{33} & t_{33}
    \end{pmatrix} \\
    = & \ \sqrt{\frac{\eta}{3}} \begin{pmatrix}
        \omega & \omega^2 & \omega^2 \\
        \omega^2 & \omega & \omega
    \end{pmatrix} \nonumber \\
    E(T) = & \ (1 - \eta) \begin{pmatrix}
        1 & 0 & 0 \\
        0 & 1 & 0 \\
        0 & 0 & 1
    \end{pmatrix} \\
    E(T)_{\fock,\fock} = & \ (1 - \eta) \begin{pmatrix}
        1 & 0 & 0 \\
        0 & 1 & 1 \\
        0 & 1 & 1
    \end{pmatrix},
\end{align}
and finally,
\begin{align}
    \brs\left(T_{\d,\fock}, E(T)_{\fock,\fock} \right) =& \\
     &  \per([T_{\d,\fock}]^\dagger T_{\d,\fock} + E(T)_{\fock,\fock}) \nonumber \\
    & - \per\left([T_{2,\fock}]^\dagger T_{2,\fock} + E(T)_{\fock,\fock}\right) \nonumber \\
    & - \per\left([T_{3,\fock}]^\dagger T_{3,\fock}+ E(T)_{\fock,\fock}\right) \nonumber \\
    & + \per(E(T)_{\fock,\fock}). \nonumber
\end{align}
Here: $T_{2,\fock} = (t_{22},t_{23},t_{23})$ and $T_{3,\fock} = (t_{32},t_{33},t_{33})$.
We will not symbolically evaluate this expression, but in the interest of providing simple test cases for future software implementations, we see numerically that $p(\d)=0.222\dots$ for $\eta=1$, $p(\d) = 0.189$ for $\eta=0.9$ and $p(\d) = 0.069444\dots$ for $\eta=0.5$.

\subsection{Using the loop Torontonian}
When performing calculations using the loop Torontonian, we need to know the matrix $O$ and the vector $\gl$ for our state.
Tools such as Strawberry Fields~\cite{killoran2019strawberry} allow for conveniently computing the real means vector, $\vec{\mu}$, and covariance matrix, $\sigma$, of a Gaussian state's Wigner function.
To use $\sigma$ and $\vec{\mu}$ to find $O$ and $\gl$, we can use functionality from The Walrus~\cite{gupt2019walrus}.
We can convert from $\sigma$ to $\cov$ by using the \texttt{Qmat} function (The Walrus uses a different ordering for $\vec{\zeta}$ and so we also apply a complex conjugate to match our definition of $\cov$), and similarly convert from $\vec{\mu}$ to $\al$ using \texttt{complex\_to\_real\_displacements}. 
These can be used to find $O = \mathbb{I} - \cov^{-1}$ and $\gl = (\cov^{-1} \al)^*$.

To show how to select the appropriate rows/columns for calculating threshold detector outcome probabilities, we will consider a 5-mode experiment, with the outcome $\d=(1,0,1,0,0)$, so $C=\{1,3\}$.
This state can be represented by a $10 \times 10$ matrix $O$, with elements $o_{jk}$, and a 10-element vector $\gl$ with elements $\gamma_j$. 
To evaluate the loop Torontonian, we form $O_{CC}$ and $\gl_C$:
\begin{align}
    O_{CC} = \begin{pmatrix}
        o_{11} & o_{13} & o_{16} & o_{18} \\
        o_{31} & o_{33} & o_{36} & o_{38} \\
        o_{61} & o_{63} & o_{66} & o_{68} \\
        o_{81} & o_{83} & o_{86} & o_{88} \\
    \end{pmatrix} \label{eq:form_O} \quad
    \gl_C = \begin{pmatrix}
        \gamma_1 \\ \gamma_3 \\ \gamma_6 \\ \gamma_8
    \end{pmatrix}
    .
\end{align}
This assumed the basis vector ordering convention specified by Eq.~21, where mode $j$ corresponds to basis vectors $j$ and $j+M$, as is used in The Walrus~\cite{gupt2019walrus}.

\section{Derivation of the characteristic function \label{deriv}}

We start from the photon number probability distribution in Eq.~6 and the definitions of the characteristic function in Eq.~7 and rearrange terms 
\begin{align}
    \chi(\vec{\phi}) & = \sum_{\fock} \exp\left(\ii \sum_{i=1}^M \phi_i n_i \right) \left|\bra{\fock} \Uhat \ket{\Phi_0}\right|^2 \\
    & = \sum_{\fock} \exp\left(\ii \sum_{i=1}^M \phi_i n_i \right) \bra{\Phi_0} \Uhat^\dagger \ket{\fock} \bra{\fock} \Uhat \ket{\Phi_0} \\
    & = \bra{\Phi_0} \Uhat^\dagger \sum_{\fock} \exp\left(\ii \sum_{i=1}^M \phi_i n_i \right) \ket{\fock} \bra{\fock} \Uhat \ket{\Phi_0}.
\end{align}
Notice that we can define an operator, $\Uhat_{\vec{\phi}}$, which acts like 
\begin{align}
    \Uhat_{\vec{\phi}} \ket{\fock} = \exp \left( \ii \sum_{i=1}^M \phi_i n_i \right) \ket{\fock},
\end{align}
as the operator given by the linear optical transformation $U_\phi$ 
\begin{align}
    U_\phi = \bigoplus_{i=1}^M \exp (\ii \phi_i).
\end{align}
If we include this in our expression above, we find 
\begin{align}
    \chi(\vec{\phi}) & = \bra{\Phi_0} \Uhat^\dagger  \Uhat_{\vec{\phi}} \sum_{\fock} \ket{\fock} \bra{\fock} \Uhat \ket{\Phi_0} \\
    & = \bra{\Phi_0} \Uhat^\dagger  \Uhat_{\vec{\phi}} \Uhat \ket{\Phi_0} \label{chi_exp},
\end{align}
where we have used the resolution of the identity to arrive at the answer.

\section{Derivation of the Bristolian \label{der_brs}}
The Unitary Bristolian has a slightly simpler derivation than the Bristolian, so we will begin by considering this case.
We start by combining Eq.~13 with Eq.~4: 

\begin{align}
p(\d) & = \\
& \left( \prod_{j=1}^M n_j!\right)^{-1}\sum_{Z \in P(\clicks)} (-1)^{|Z|} \per([U^\dagger U_{\vec{x}(V,Z)} U]_{\fock, \fock}). \nonumber
\end{align}
$U_{\vec{x}(V,Z)}$ is formed by defining $U_{\vec{x}}$ with $x_j=0$ if $j \in V$ or $j \in Z$, and $x_j=1$ otherwise.
Since $U_{\vec{x}(V,Z)}$ contains only zeros on the rows and columns given by the elements of $V$ and $Z$, this is equivalent to deleting the rows and columns of $U$ and $U^\dagger$ respectively according to the elements of $V$ and $Z$. 
Therefore, we can write the sum above as 
\begin{align}
     \sum_{Y \in P(\clicks)} (-1)^{|C| - |Y|} \per\left([(U_Y)^\dagger U_Y]_{\fock,\fock}\right) \label{brs2}.
\end{align}
Here, we sum over the modes which are marginalised, instead of summing over the modes being projected into the vacuum state. 
This corresponds to using the substitution $Z$ for $Y = \clicks \setminus Z$, and noticing that the sum over $Y = \clicks \setminus Z:\ Z \in P(\clicks)$ is the same as the sum over $Z \in P(\clicks)$.
We use $U_Y$ to denote selecting only the rows of $U$ according to the elements in the set $Y$.
Note that since we delete rows when all their elements are set to all zeros, the permanent when $Y$ is the empty set should be zero, as it corresponds to the permanent of an all zeros matrix, rather an empty matrix.

Because Eq.~\eqref{brs2} contains an inclusion/exclusion formula, like that of Ryser's permanent formula~\cite{Ryser1963}, it could be viewed as a 3-dimensional permanent, similar to those which appear elsewhere in quantum photonics~\cite{tichy2015sampling, rudolph2021perhaps}.
However, we note that to compute the probability for input state $\ket{\fock}$, interferometer transformation $U$ and threshold detector pattern $\d$, it is sufficient to know only the rows of $U$ which correspond to nonzero elements of $\d$ and columns of $U$ given by $\fock$.
Therefore the input to this problem is a matrix, so we chose to write this probability in terms of a new matrix function, the \textit{Unitary Bristolian} 
\begin{align}
    p(\vec{d}) = \frac{\ubrs(U_{\d,\fock})}{\prod_{j=1}^M n_j !}.
\end{align}

We construct $U_{\d,\fock}$ from $U$ as described under Eq.~12.
The Unitary Bristolian, $\ubrs$, is a matrix function which acts on some $m \times n$ matrix, $A$:
\begin{align}
    \ubrs(A) = \sum_{Y \in P([m])} (-1)^{m - |Y|} \per([A_Y]^\dagger A_Y),
\end{align}
which is the form this is reported in Eq.~18.

Now, we are ready to derive the more general formula for the Bristolian.
Unlike the permanent, the Bristolian can be generalised for calculating marginal detection probabilities. 
Consider that we wish to calculate the probability of observing detector clicks for modes in $C$, vacuum in modes $V$, and marginalise over modes in $B$, so the union of $\clicks$, $\vacs$ and $B$ is $[M]$.
We can write this probability as 
\begin{align}
    &p(\d_\clicks = \vec{1}, \d_V = \vec{0}) = \label{mprob} \\
    &\left( \prod_{i=1}^M n_i!\right)^{-1} \sum_{Y \in P(\clicks)} (-1)^{|\clicks| - |Y|} \per\left([(U_{Y\cup B})^\dagger U_{Y\cup B}]_{\fock,\fock}\right). \nonumber
\end{align}
Here $U_{Y \cup B}$ is constructed by selecting rows of $U$ according to the elements of $Y$ and $B$. 
We could also write the summation in Eq.~\eqref{mprob} as 
\begin{align}
    \sum_{Y \in P(\clicks)} (-1)^{|\clicks| - |Y|} 
    \per\left[ 
\begin{pmatrix}U_{Y,\fock} \\  U_{B,\fock} \end{pmatrix}^\dagger \begin{pmatrix}U_{Y,\fock} \\  U_{B,\fock} \end{pmatrix}
    \right],
    \label{mprob2}
\end{align}
where $\begin{pmatrix}U_{Y,\fock} \\  U_{B,\fock} \end{pmatrix}$ is the augmented $(|Y| + |B|) \times N$ matrix formed by stacking the matrices $U_{Y,\fock}$ and $U_{B,\fock}$, where $N=\sum_j n_j$.
The notation $U_{\clicks, \fock}$ is used to show that we take rows of $U$ according to the set $\clicks$ and repeat the columns of $U$ according to $\fock$, and equivalently for $U_{B,\fock}$.

A counter-intuitive feature of this formula is that it depends on matrix elements of the linear transformation that are ignored by our measurements.
In an experiment, changing the elements of $U_B$ should have no impact on $p(\d_\clicks = \vec{1}, \d_V = \vec{0})$.
Following this argument, we propose that we only need to know $U_C$, and we can construct $U_B$ by performing a unitary dilation.
This is particularly helpful for lossy experiments, where we typically do not have an understanding of the full unitary transformation acting on the both the experiment's (lossy) modes and the loss modes of its environment.

Any open quantum dynamics can be expressed as unitary evolution on of a larger system via unitary dilation.
For a non-unitary transformation given by a rectangular matrix, $T$, with singular values all $\leq 1$, we can write the unitary dilation 
\begin{align}
    U(T) = \begin{pmatrix} T & \left(\mathbb{I} - T T^\dagger\right)^{1/2} \\
    \left(\mathbb{I} - T^\dagger T\right)^{1/2} & -T^\dagger
    \end{pmatrix}.
    \label{dilation}
\end{align}

With this, we are now ready to write down the probability for the general case of a Fock state input, linear optical experiment.
We consider that we have $M_\text{in}$ input modes, initialised in the state $\fock$. 
These propagate through a nonunitary linear transformation, $T$, before being detected by $M_\text{out}$ threshold detectors, which give an outcome $\d$, which is a length-$M_\text{out}$ bit-string. 
So $T$ is given by an $M_\text{out} \times M_\text{in}$ matrix. 

First, we dilate $T$ according to Eq.~\eqref{dilation}, giving an $(M_\text{in} + M_\text{out})$-dimension unitary matrix.
In this construction, $U_\clicks=T$ and $U_{B}$ is given by $\left(\mathbb{I} - T^\dagger T \right)^{1/2}$.
We also notice that we can explicitly write out the multiplication of the augmented matrices:
\begin{align}
    \begin{pmatrix}
        A \\ R
    \end{pmatrix}^\dagger 
    \begin{pmatrix}
        A \\ R
    \end{pmatrix} = 
    A^\dagger A  + R^\dagger R.
\end{align}
Combining these observations with Eq.~\eqref{mprob} and Eq.~\eqref{mprob2}, we can write down the probability of measuring a click pattern $\vec{d}$ on a Fock state $\fock$ evolving through a nonunitary transformation $T$ as 
\begin{align}
    p(\d) = \frac{\brs\left(T_{\d,\fock}, E(T)_{\fock,\fock} \right)}{\prod_{j=1}^{M} n_j ! }.
\end{align}
Where we have introduced a new matrix function, the \textit{Bristolian}, $\brs$ 
\begin{multline}
    \brs\left(A, E \right) = \\
    \sum_{Y \in P([m])} (-1)^{m - |Y|} 
    \per \left(
    [A_Y]^\dagger A_Y + E\right),
\end{multline}
where $m = \sum_j d_j = |C|$, is the total number of clicks.
We also define the matrix 
\begin{align}
    E(T) = \mathbb{I} - T^\dagger T.
\end{align}
To arrive at this equation, we have used
\begin{align}
    R(T)_{:,\fock} &= \left[\left( \mathbb{I} - T^\dagger T \right)^{1/2}\right]_{:,\fock} \\
    [R(T)_{:,\fock}]^\dagger R(T)_{:,\fock} &= E(T)_{\fock,\fock},
\end{align}
where the subscript notation $:,\fock$ means that we select the columns according to $\fock$.

\section{Unitary Bristolian reduction to the permanent for clicks equal to photons \label{brs_per}}
When the number of input photons is equal to the number of threshold detector clicks, the Unitary Bristolian has a square matrix as an input.
To see $N$ clicks for an $N$ photon input state, we know that each threshold detector must have detected exactly 1 photon, and so this event could also be modelled by using photon number projectors instead of the click measurement operator, and we can use Eq.~12 to calculate its amplitude.
Therefore, we expect that the Bristolian of a square matrix should reduce to the absolute square of the permanent of the same matrix.

For a square $N \times N$ matrix, $A$, with elements $a_{jk}$, we can write the Unitary Bristolian as 
\begin{align}
    \ubrs(A) = &  \sum_{Y \in P([N])} (-1)^{N - |Y|} \per([A_Y]^\dagger A_Y).
\end{align}
Then we can expand the permanent using Ryser's formula 
\begin{multline}
     \per([A_Y]^\dagger A_Y) = \\ \sum_{Z \in P([N])} (-1)^{N - |Z|} \prod_{i=1}^N \sum_{j \in Z} \left[(A_Y)^\dagger A_Y \right]_{ij}
\end{multline}
\begin{align}
    = \sum_{Z \in P([N])} (-1)^{N - |Z|} \prod_{i=1}^N \sum_{j \in Z} \sum_{k \in Y} a^*_{ki} a_{kj}.
\end{align}
This gives 
\begin{align}
    \ubrs(A) = & 
    \sum_{Y, Z \in P([N])} (-1)^{|Y| + |Z|} \prod_{i=1}^N \sum_{\substack{j \in Z \\ k \in Y}} a^*_{ki} a_{kj},
\end{align}
which is the Ryser-style formula for the 3-tensor permanent, as defined in Ref.~\cite{tichy2015sampling}  
\begin{align}
   \ubrs(A) = \per(B) = \sum_{\sigma,\rho \in S_N} \prod_{i=1}^N b_{i \sigma(i) \rho(i)},
\end{align}
with $\sigma$ and $\rho$ being elements of the permutation group $S_N$. 
However, this 3-tensor, $B$, with elements $b_{ijk} = a^*_{ki} a_{kj}$ is very structured.
This kind of structure is discussed in Ref.~\cite{tichy2015sampling}, where it is shown that it allows us to factor this expression into 
\begin{align}
    \ubrs(A) = & \per(A^*) \per(A) = \per(A)^* \per(A) \\
    = &\ |\per(A)|^2.
\end{align}

\section{Computing the Bristolian as a sum over Fock state probabilities  \label{app:brs_using_pers}}
In the absence of our expression for the Bristolian in Eq.~15, the only known way in the literature to compute probabilities for Fock states measured with threshold detectors is to consider all possible events which could lead to the witnessed outcome, and sum all their corresponding probabilities.

To provide an example for how the complexity of this method compares to the Bristolian, we consider an $M$ mode experiment with uniform transmission, $\eta$, $N$ input photons and $n$ clicks (with $n \leq N$).
For interferometers with imbalanced losses, we must consider losses at both the input and output of the interferometer.
However, here we are considering balanced loss which allows us to consider that any photon loss occurs before the photons reach the interferometer, as balanced loss commutes with linear optics.
Because any number of photons between $n$ and $N$ can be transmitted through this loss channel and lead to an $n$ click event, we must consider all $\sum_{j=n}^N \binom{N}{j}$ possible configurations for how these photons could have been transmitted.
For each input configuration, we then need to consider all the ways that these photons can bunch within the detectors.
We know that at least 1 photon must arrive in each detector, which leaves $j-n$ photons left, which can be configured in any arrangement.
There are $\binom{X+Y-1}{Y}$ ways of arranging $Y$ photons into $X$ modes, and here we have $j-n$ photons which can be arrive in $n$ modes, meaning we need to calculate $\binom{j-n+n-1}{j-n}=\binom{j-1}{j-n}$ permanents for each input configuration.
This results in a total of $\sum_{j=n}^N \binom{N}{j} \binom{j-1}{j-n}$ permanents.

Each permanent for a $j$ photon configuration has a time complexity of $j2^j$, giving an overall time complexity which is lower bounded by $\binom{N}{n} n 2^n$.
By comparison, the Bristolian complexity of $n2^{2n}$ provides a superexponential speedup when $N$ and $N-n$ are large.

\section{Derivation of the loop Torontonian \label{ltor_derivation}}
We start with some definitions. 
For a Gaussian state with complex Husimi covariance matrix $\cov$ and complex vector of means $\alpha$ we define 
\begin{align}
\om &= \mathbb{I} - \cov^{-1}, \label{eq:defOg} \\
\gl &= (\cov^{-1} \al)^*, \label{eq:def1g}
\end{align}
which uniquely specify the \emph{photon number} statistics of the Gaussian state~\cite{quesada2019simulating} via loop Hafnians~\cite{bjorklund2019faster}. Given a photon number outcome $\vec{n} = (n_1,\ldots,n_M)$ its probability is given by
\begin{align}
p(\vec{n}) = p(\vec{0}) \lhaf(X \om_{\fock,\fock}, \gl_{\fock}),
\end{align}
where $p(\vec{0})$ is the vacuum probability, $X = \left[\begin{smallmatrix} 0 & \mathbb{I} \\ \mathbb{I} &  0 \end{smallmatrix}\right] $ and $\om_{\fock,\fock}$ and $\gl_{\fock}$ are submatrices of $\om$ and $\gl$, found by using $n_j$ repetitions of the rows/columns corresponding to mode $j$ (recalling that each mode $j$ corresponds to 2 rows/columns of $O$).

We will now show that the threshold probabilities can also be written in terms of the quantities defined in Eq.~\eqref{eq:defOg} and Eq.~\eqref{eq:def1g}.

By ordering our basis vectors such that modes which see a click ($\clicks$) are arranged to be before modes which see vacuum ($\vacs$), the matrix $O$ can be written in block form 
\begin{align}
\om = \begin{pmatrix}
    \om_{\clicks\clicks}  & \om_{\clicks\vacs} \\
    \om_{\vacs\clicks} & \om_{\vacs\vacs}
    \end{pmatrix},
\end{align}
then, using Schur complements, we can see that 
\begin{align}\label{eq:schurblock}
\om_{\clicks\clicks} = \mathbb{I} - \left[\cov_{\clicks\clicks} - \cov_{\clicks\vacs} [\cov_{\vacs\vacs}]^{-1} \cov_{\vacs\clicks} \right]^{-1}.
\end{align}
We are now ready to investigate threshold probabilities. 
We start with Eq.~4 which we write as
\begin{align}
p(\d) = p(\vec{0}) \sum_{Z \in P(\clicks)} (-1)^{|Z|} \frac{p(\vec{m}_Z = 0, \vec{m}_V=0)}{p(\vec{0})}.
\end{align}
For a given $Z$, the term inside the sum can be written as in Eq.~23.
Note that the argument inside the exponential in said equation can be rewritten as
\begin{align}
\al_W^\dagger [\cov_{WW}]^{-1} \al_W =& \gl_W^t \cov_{WW} \gl_W^* \\
& + \gl_W^t \cov_{W Y} \gl_Y^* + \gl_Y^t \cov_{YW} \gl_W^*  \nonumber \\
&+ \gl_Y^t \cov_{YW} [\cov_{WW}]^{-1} \cov_{WY} \gl_Y^*, \nonumber 
\end{align}
where $W$ denotes the union of sets $Z$ and $V$, and $Y$ denotes the modes not included in $W$.
Here we have used that $\al_W = \cov_{WW} \gl_W^* + \cov_{WY} \gl_Y^*$, $\cov_{WW}^\dagger = \cov_{WW}$ and $\cov_{WY}^\dagger = \cov_{YW}$.
We can similarly write 
the argument of the exponential in $p(\vec{0})$ as
\begin{align}
\al^\dagger [\cov]^{-1} \al = \gl^t \cov \gl^* =& \gl_W^t \cov_{WW} \gl_W^* + \gl_W^t \cov_{WY} \gl_Y^* \\&+ \gl_Y^t \cov_{YW} \gl_W^*  + \gl_Y^t \cov_{YY} \gl_Y^*. \nonumber 
\end{align}
With these two expressions we can then write 
\begin{multline}
\frac{p(\vec{m}_W = 0)}{p(\vec{0})} = 
\sqrt{\frac{\det(\cov)}{\det(\cov_{WW})}}  \\
\times
\exp\left[ \tfrac12 \gl_Y^t [\cov_{YY} - \cov_{YW} [\cov_{WW}]^{-1} \cov_{WY}] \gl_Y^* \right] \nonumber
\end{multline}
\begin{align}
= \frac{\exp\left[ \tfrac12 \gl_Y^t [\mathbb{I} - \om_{YY}]^{-1} \gl_Y^* \right]}{\sqrt{\det(\mathbb{I} - \om_{YY})}}.
\end{align}
In the last equation we used the result in Eq.~\eqref{eq:schurblock}, together with factorising the determinant
\begin{multline}
    \det(\cov) = \\
    \det(\cov_{WW}){\det(\cov_{YY} - \cov_{YW} [\cov_{WW}]^{-1} \cov_{WY})},
\end{multline} 
to show
\begin{align}
\frac{\det(\cov)}{\det(\cov_{WW})} =& \det(\cov_{YY} - \cov_{YW} [\cov_{WW}]^{-1} \cov_{WY}) \\
=& \det\left([\mathbb{I} - O_{YY}]^{-1} \right)\\
=& \frac{1}{\det(\mathbb{I} - \om_{YY})}.
\end{align}
This corresponds to the identity $\det(\cov_{WW}) = \det(\cov)\det([\cov^{-1}]_{YY})$ from Ref.~\cite{shi2022effect}.
With these observations we can write the sought after probability as 
\begin{align}
p(\d) = p(\vec{0}) \ltor\left( \om_{CC}, \gl_C \right),
\end{align}
where $C$ is the set of modes with threshold detector clicks, and $\ltor$ stands for the loop Torontonian defined as 
\begin{multline}
\ltor\left( \om, \gl \right) = \\
\sum_{Y \in P([m])} (-1)^{m-|Y|} \frac{\exp\left[ \tfrac12 \gl_Y^t [\mathbb{I} - \om_{YY}]^{-1} \gl_Y^* \right]}{\sqrt{\det(\mathbb{I} - \om_{YY})}}
\end{multline}
for a $2m \times 2m$ matrix $O$, and a $2m$-length vector $\gl$.

In arriving at this formula, we have swapped the summation over $Z$ for a summation over $Y$, as we did in Eq.~\eqref{brs2}.
This is the form that the equation appears in the main text.

\section{Generating the loop Hafnian from the loop Torontonian \label{lhaf}}
Just like the Torontonian~\cite{quesada2018gaussian}, the loop Torontonian has an interesting interpretation of being a generating function for photon number probabilities of Gaussian states (now with nonzero displacement).

To this end, recall that a threshold probability can be obtained as a sum (coarse graining) of many photon number events.
For any $\d$, we can write
\begin{align}
\frac{p(\d)}{p(\vec{0})}	 =& \frac{p(\vec{n} = \d)}{p(\vec{0})} + \sum_{\vec{k} \in \mathcal{\clicks}_{\d}}  \frac{p(\vec{n} = \d+\vec{k})}{p(\vec{0})} \label{eq:prob_sum}\\
\ltor\left( \om_{CC}, \gl_{C} \right) =&  
\lhaf(X \om_{\vec{d}, \vec{d}}, \gl_{\vec{d}}) 
\label{eq:lhaf_sum} \\&+  \sum_{\vec{k} \in \mathcal{\clicks}_{\d}}  \lhaf(X \om_{(\vec{d}+\vec{k}),(\vec{d}+\vec{k})}, \gl_{\vec{d}+\vec{k}}),\nonumber
\end{align}
where $\mathcal{\clicks}_{\vec{d}}$ is the set of all vectors of integers that have zero in the positions where $\vec{d}$ has zero and strictly positive integers in all the other positions. Note that any element $\vec{k}$ in the set satisfies $K = \sum_{i=1}^M k_i>0$.
We now recall that loop hafnians satisfy the following scaling property~\cite{bjorklund2019faster} 
\begin{align}
\lhaf((\eta {A})_{\fock,\fock},(\sqrt{\eta}{\gamma})_{\vec{n}}) = \eta^{|N|} \lhaf({A}_{\fock,\fock},{\gamma}_{\vec{n}}),
\end{align}
with $N = \sum_j n_j$. 
This allows us to write 
\begin{multline}
\ltor\left( \eta \om_{CC}, \sqrt{\eta} \gl_{C} \right) = \eta^{|C|}\lhaf(X \om_{\d,\d}, \gl_{\d}) \\+ 
\sum_{\vec{k} \in \mathcal{\clicks}_{\d}} \eta^{|C|+K} \lhaf(X \om_{(\vec{d}+\vec{k}),(\vec{d} +\vec{k})}, \gl_{(\vec{d}+\vec{k})}).
\end{multline}
If we set $\ell = |C|$ we can derive a formula for the loop hafnian
\begin{align}
\lhaf(X \om_{\vec{d},\vec{d}}, \gl_{\vec{d}}) =& \left.\frac{1}{\ell!} \frac{d^\ell}{d \eta ^{\ell}} \ltor\left( \eta \om_{\vec{d}, \vec{d}}, \sqrt{\eta} \gl_{\vec{d}} \right) \right|_{\eta = 0}  \\
= & \sum_{Y \in P([m])} (-1)^{m-|Y|} f\left(O_{YY}, \gl_{Y}\right)
\end{align}
where we define 
\begin{equation}
    f(O,\gl) = \left. \frac{1}{\ell!} \frac{d^\ell}{d \eta^\ell} q(O,\gl) \right|_{\eta=0}
\label{f_exp}
\end{equation}
\begin{equation}
    q(O, \gl) = \frac{\exp\left[ \tfrac12 \sqrt{\eta }\gl^t [\mathbb{I} - \eta \om]^{-1}  \sqrt{\eta} \gl^* \right]}{\sqrt{\det(\mathbb{I} - \eta \om)}}.
\end{equation}
One way of interpreting $f(O,\gl)$ 
is that it selects the $\ell$th coefficient of the polynomial expansion of $q$. 
Therefore, we do not require knowledge of $q(O,\gl)$ beyond order-$\ell$.
Using the Mercator series, we expand the denominator in Eq.~\eqref{f_exp} as 
\begin{align}
    \det(\mathbb{I} - \eta O)^{-1/2} = \exp \left[ 
    \sum_{k=1}^{\infty}
    \frac{\tr\left(\eta^k O ^ k \right)}{2k}
    \right],
\end{align}
so we can combine the denominator into the exponent.
We also notice that we can use a binomial expansion, $(1-x)^{-1} = \sum_{j=0}^\infty x^j$,  to substitute  $
    [\mathbb{I} - \eta O]^{-1} = 
    \sum_{k=1}^\infty \left( \eta O \right)^{k-1}
$.
This gives us an expression 
\begin{multline}
    q(O,\gl) = \\
    \exp\left(\sum_{k=1}^{\infty} \left[
    \frac{\tr\left(O^k \right)}{2k} + 
    \frac{\gl^t \left( O^{k-1} \right) \gl^*}{2} 
    \right] \eta^k
    \right),
\end{multline}
which we Taylor-expand up to order $\ell$ to obtain
\begin{multline}
    q(O,\gl) = \\
    \sum_{j=0}^{\ell} \frac{1}{j!}
    \left(
    \sum_{k=1}^{\ell} \left[
    \frac{\tr\left(O^k \right)}{2k} + 
    \frac{\gl^t \left(O^{k-1} X \right) \gl}{2} 
    \right] \eta^k
    \right)^j.
\end{multline}
We have used that $\gl^*=X \gl$.
This gives us the exact form of the trace formula algorithm for the loop Hafnian as presented in Ref.~\cite{bjorklund2019faster} (bar some typos in the referenced paper).
It is interesting that this derivation uses various arguments which rely specifically on $O$ and $\gl$ being formed by a Gaussian state covariance matrix (in particular Eq.~\eqref{eq:prob_sum} and Eq.~\eqref{eq:lhaf_sum}),
however the trace formula algorithm that we are able to derive here is applicable to arbitrary symmetric matrices.

\begin{figure}[!h]
	    \includegraphics[width=0.7\columnwidth]{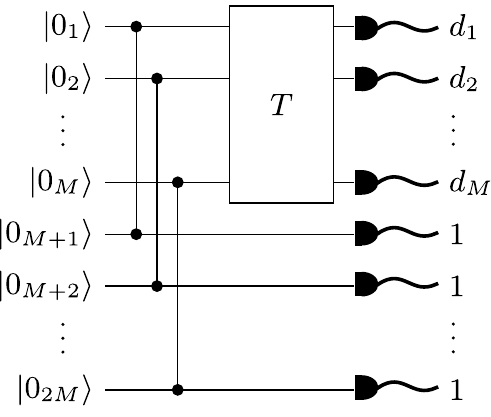}	\caption{\label{fig:scatter} Scattershot construction. Gates between modes, symbolised by vertical lines with dots at the ends, show weak two-mode squeezing. The two-mode squeezing operator between modes $i$ and $j$ is given $\hat{S}_{i,j}(t) = \exp\left( r \left[\hat{a}_i^\dagger \hat{a}_j^\dagger - \hat{a}_i \hat{a}_j\right] \right)$.}
\end{figure}
\section{Connecting the Bristolian and the Torontonian}\label{app:bris_to_tor}

In this Appendix we provide a formal link between the Bristolian and the Torontonian. Concretely, we show that the Bristolian associated with the threshold detection of a multimode Fock state with single photon or vacuum inputs can be evaluated as a certain limit of a Torontonian.

The starting point of our derivation is the scattershot Boson sampling construction shown in Fig.~\ref{fig:scatter} and introduced in Ref.~\cite{lund2014boson}.
The diagram represents $M$ two-mode squeezed vacuum states where the first half of the modes (the heralded modes) are sent into an interferometer with transmission matrix $T$, and the second half of the modes (the heralding modes) are sent into threshold detectors.

For each individual two-mode squeezed vacuum with squeezing parameter $t$, it is straightforward to show that conditioned on a click in the heralding mode, the state of the heralded mode collapses to~\cite{bourassa2021fast}
\begin{align}
\rho_{\checkmark} = \frac{1}{1+\bar{n}} \pr{1} + \sum_{m=2}^\infty \frac{1}{\bar{n}} \left(\frac{\bar{n}}{1+\bar{n}} \right)^m \pr{m},
\end{align}
where $\bar{n} = \sinh^2 t$ is the mean photon number of either mode of the two-mode squeezed vacuum. The probability of heralding the state $\rho_{\checkmark}$ is given by
\begin{align}
p_{\checkmark} = \frac{\bar{n}}{1+\bar{n}} = \varepsilon^2\text{ with } \varepsilon = \tanh t .
\end{align}
Note that as $\varepsilon \to 0$ the fidelity between $\rho_{\checkmark}$ and a single photon Fock state approaches one, but at the same time the probability of heralding the state approaches zero. 

We can now study the covariance matrix $\cov$ of the Gaussian circuit in Fig.~\ref{fig:scatter}. After some algebra, it can be shown that the matrix $O = \mathbb{I} - \cov^{-1}$ dictating the threshold probabilities of the Gaussian state is given by
\begin{align}
O(\varepsilon) = \varepsilon \begin{pmatrix}
	0 & 0 & 0 & T \\
	0 & \varepsilon E(T)^*
	& T^t & 0 \\
	0 & T^* & 0  & 0\\
	T^\dagger &  0 & 0 & \varepsilon E(T)
	\end{pmatrix}.
\end{align} 
We now want to calculate the probability that a subset of the first $M$ modes clicks conditioned on all the modes on the second half clicking. We specify the modes that clicked in the first half by $\vec{d}$ and then can write the click pattern for the $2M$ modes to be $\vec{e} = \vec{d} \oplus \vec{1}$ where, recall, $\vec{1}$ is the all ones vector of length $M$. We can now write the conditional probability as
\begin{align}
p(\vec{d}|\vec{1}_M) = \frac{p(\vec{d} \oplus \vec{1})}{p_\checkmark^M} = \frac{p(\vec{e})}{p_\checkmark^M}.
\end{align}
For the probability in the numerator of the right-hand side in the last equation we can use Eq.~25. 
The vacuum probability of the $2M$-mode state is given by
\begin{align}
p(\vec{0}) = \sqrt{\det(\mathbb{I} - O(\varepsilon))} = (1-\varepsilon^2)^{M}.
\end{align}
We can now write
\begin{align}
p(\vec{d}|\vec{1}_M) = \frac{(1-\varepsilon^2)^{M} }{ \varepsilon^{2M}} \tor(O(\varepsilon)_{CC})
\end{align}
where $C$ is the union of the $\{M+1,\ldots,2M\}$ heralding modes and the modes that have a one in the vector $\vec{d}$, i.e., the modes that click in the first half.

As explained at the beginning of this section, in the limit where $\varepsilon \to 0$ we know that the conditional state of the input heralded modes becomes a product of single photons in each mode, and thus in this same limit we can write
\begin{multline}
\brs\left(T_{\vec{d},\vec{1}}, E(T)_{\vec{1},\vec{1}}\right) = \\ \lim_{\varepsilon \to 0 }  (\varepsilon^{-2} - 1)^M  \tor\left( O(\varepsilon)_{CC} \right). 
\end{multline}
Note that we have so far only considered the case where single photons are input in all the modes of the interferometer. The more general case where vacuum is fed into some of the modes can be dealt with by applying a loss channel with zero-transmission to the relevant modes. This is equivalent to setting to zero the columns of $T$ where vacuum is fed. Using this argument we find that for cases where $\fock$ is a bistring, i.e., we only allow single photons inputs, and setting $N = \sum_{i} \fock_i$ we can write
\begin{multline}
\brs\left(T_{\vec{d},\fock}, E(T)_{\fock,\fock} \right) = \\
\lim_{\varepsilon \to 0 }  (\varepsilon^{-2} - 1)^{N}  \tor\left( O(\varepsilon)_{CC} \right), 
\end{multline}
where now $C$ is the union of the labels of the modes in which single photons where input into the interferometers and the labels of the modes in which clicks are registered. It is interesting to consider that one could potentially derive the form of the Bristolian in terms of sums of permanents by using the connection between permanents and determinants provided by the MacMahon Master theorem~\cite{macmahon2001combinatory}.

Finally, note that one can also write circuits to herald multi-photon Fock states using only threshold detectors as shown in Appendix D of Ref.~\cite{bourassa2021fast}.

\section{Time complexities \label{app:time_comp}}
To calculate a marginal vacuum probability for a Fock state evolved through a linear interferometer, we can compute a permanent, as given in Eq.~13
If we are detecting vacuum in modes given by $\vacs$ and marginalising over all other modes, given by $B$, then the matrix $[U^\dagger U_{\vec{x}} U]_{\fock,\fock}$ has rank $\leq |B|$, 
so its permanent can be computed in time $\bigO(N^{\bigO(|B|)})$ using the algorithm introduced in Sec. III of Ref.~\cite{barvinok1996two}.
If we are marginalising over many modes, and therefore $|\vacs|$ is small, we can instead consider using the algorithm introduced in Sec. IV. A of Ref.~\cite{ivanov2020complexity} to compute the permanent in time $\bigO(N^{2|V|+1})$.

For a general permanent of an $n \times n$ matrix, Ryser's algorithm~\cite{Ryser1963} has the best known complexity of $\bigO(n2^n)$.
For computing the Bristolian, to calculate the permanents inside the summation in Eq.~15, we may sometimes be able to use the faster algorithms above, however the dominant complexity for this formula would still come from computing the intermediate cases, when neither of the faster algorithms are applicable, where Ryser's algorithm may be the fastest option.
This is upper bounded by $\bigO(n2^n)$ and there are $\bigO(2^m)$ terms in the sum.
Therefore, a Bristolian of an $m \times n$ matrix $A$, and $n \times n$ matrix $E$ has a time complexity of $\bigO(n2^{m+n})$. 

We do not claim that these complexities are optimal.
The structure of this matrix function may be exploited to reduce the complexity, for example by using methods similar to those for low rank permanents~\cite{barvinok1996two}, exploiting recursion~\cite{kaposi2021polynomial} and using Laplace expansions~\cite{clifford2018classical}.
However, we leave it as an open problem to find faster algorithms for the Bristolian.

For the loop Torontonian, we find a comparable complexity to the original algorithm for the Torontonian, with complexity $\bigO(m^3 2^m)$ for a $2m \times 2m$ matrix $O$ and $2m$-length vector $\gl$.
In each step, we must compute a matrix inverse and a matrix determinant, both having $m^3$-time algorithms.
However, these steps can make use of the Cholesky decomposition of $O$, so we can improve the polynomial prefactor, following the methods described in Ref.~\cite{kaposi2021polynomial}.

Both of these methods see a quadratic penalty as compared to the fastest methods for calculating photon number probabilities of pure states.
This can be understood as being caused by the threshold detection operators in Eq.~1b having high rank, whereas the photon number operators are rank-1 projectors.
For Gaussian state calculations, we also see a quadratic cost for calculating photon number probabilities when the state is mixed~\cite{kruse2019detailed, quesada2019simulating}, so we can also understand this quadratic penalty as being a result of introducing mixture into the projected state.
This differs from the case of sampling, where it was shown that sampling threshold detector clicks can be simulated with the same complexity as single photon measurements~\cite{bulmer2021boundary}.
If we accept approximate expressions, accurate to additive error, we can efficiently compute the probabilities presented in this work using Monte-Carlo phase space methods~\cite{drummond2022simulating}.
However, for events with small probability, these methods can quickly become impractical due to a large relative error.

\section{Accuracy improvements of a threshold detection model over a single photon projection approximation \label{app:accuracy} }  
\begin{figure}[t]
    \centering
    \includegraphics[width=0.97\columnwidth]{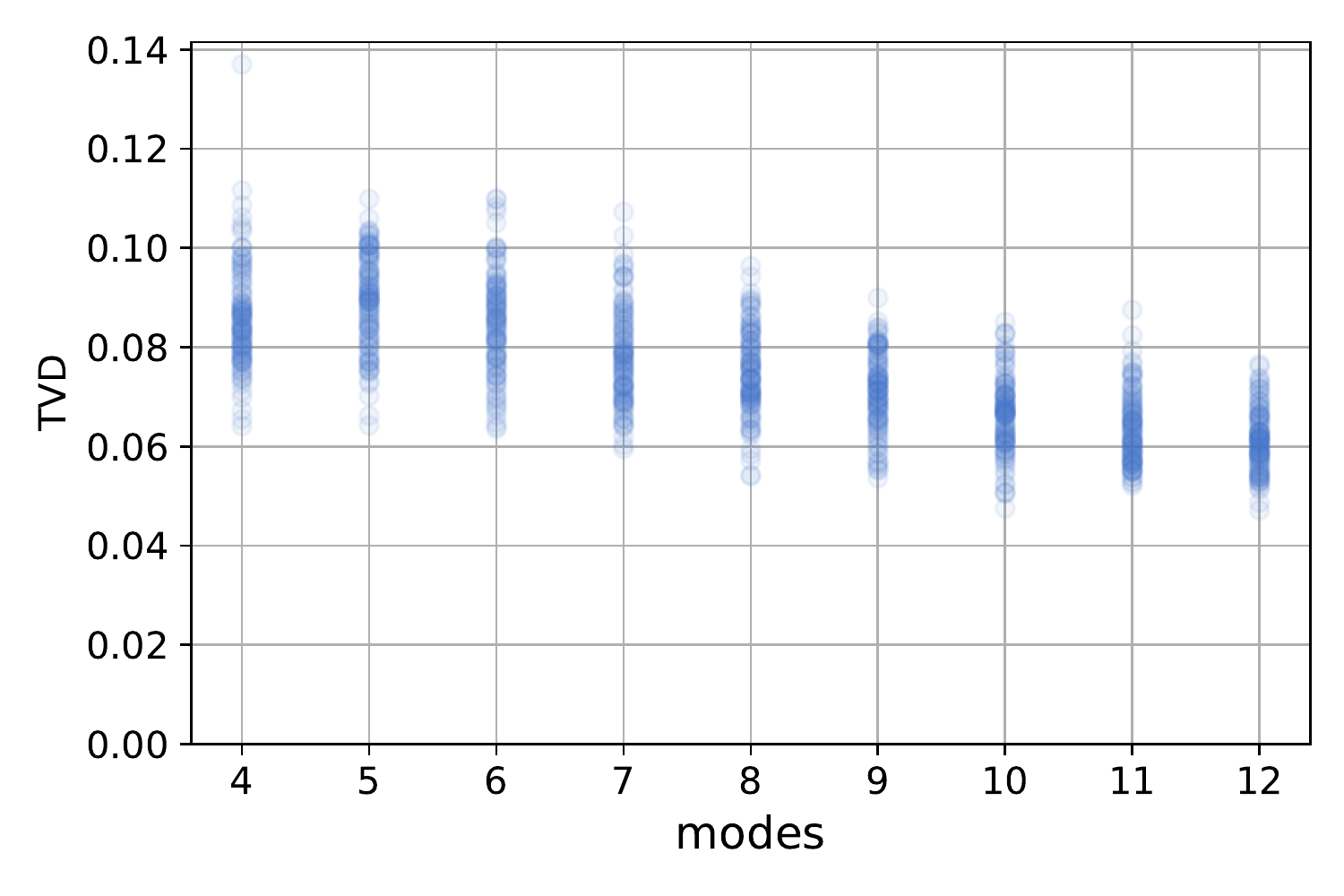}
    \caption{Total variation distance (TVD) between a simulation of a threshold detection based, 4 photon, Fock state boson sampling experiment and an approximate model which uses Fock state projection.
    For each number of modes, 100 Haar random unitary matrices are sampled and the full probability distribution is calculated.
    Threshold detection calculations are performed using the Bristolian, Fock state projection calculation uses a formula based on matrix permanents, as discussed in Appendix~\ref{app:brs_using_pers}.}
    \label{fig:fock_bs_tvd}
\end{figure}

To provide a quantitative demonstration of the accuracy improvements of an exact model of threshold detection versus the typical approximation of using single photon projection, we calculate the full probability distribution for a 4 photon Fock state boson sampling experiment~\cite{aaronson2011computational} using the Bristolian with the number of modes ranging from 4 to 12, choosing a transitivity of $\eta = 0.6$.
We also find the probability distribution given by 0 or 1 photon Fock state projective measurement.
We calculate the total variation distance (TVD) between these distributions for 100 different Haar random linear optical interferometers for each number of modes.
The results, shown in Fig.~\ref{fig:fock_bs_tvd}, show that even as we approach the $M = N^2$ regime, we do not see a convergence between these two distributions.
In particular, we see that the TVD for all the experiments lies typically within the range of 5\%-12\%.
This highlights the importance of using the correct mathematical description of the experiment in order to best understand the results.

\begin{figure}[t]
    \centering
    \includegraphics[width=\columnwidth]{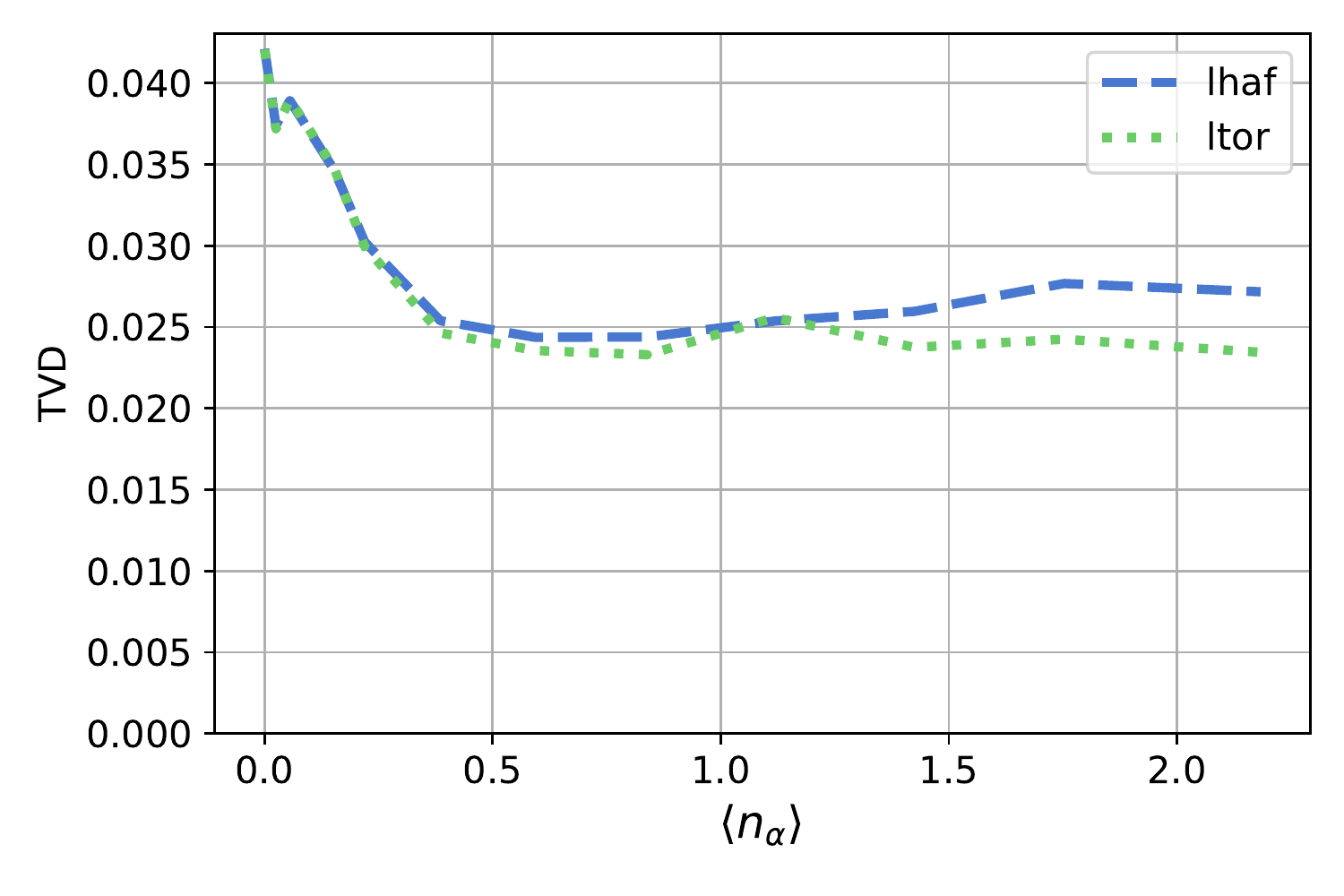}
    \caption{Total variation distance (TVD) of 2 photon displaced Gaussian boson sampling distributions from the experiment reported in Ref.~\cite{thekkadath2022experimental} against a threshold detector model, using the loop Torontonian ($\ltor$, dotted line), and a Fock state projection model, using the loop Hafnian ($\lhaf$, dashed line).
    This is calculated for different mean photon numbers of the input coherent state $\langle n_\alpha \rangle$.}
    \label{fig:dispGBS}
\end{figure}

We also performed an analysis of data from a recent experiment of displaced Gaussian boson sampling~\cite{thekkadath2022experimental}, which used a two-mode squeezed vacuum and a single coherent state as input states.
Here, we look at the 2 photon probability distribution as estimated using the experiment for different levels of displacement, as labelled by the mean photon number of the input coherent state, $\langle n_\alpha \rangle$, in Fig.~\ref{fig:dispGBS}.
For each probability distribution, we compare the TVD to a model which uses threshold detection, using the loop Torontonian, and a model which assumes Fock state projections, using the loop Hafnian.
The experiment of Ref.~\cite{thekkadath2022experimental} uses threshold detection, and we see that the model which uses loop Torontonians provides a more accurate model of the experiment.
For the largest displacements, when looking at the ratio of the TVD, the loop Torontonian gives a probability distribution which is 16\% closer to the experimental data.

\section{Software implementation \label{app:code}}
For an efficient and parallelisable implementation of the Bristolian and loop Torontonian, we use just-in-time compilation provided by \texttt{Numba}~\cite{lam2015numba}.
Our code is available in the open-source Python package \texttt{The Walrus}~\cite{gupt2019walrus} (\url{https://github.com/XanaduAI/thewalrus}) in releases from 0.19.0 onward, and were contributed in pull requests \href{https://github.com/XanaduAI/thewalrus/pull/316}{(316)} and \href{https://github.com/XanaduAI/thewalrus/pull/317}{(317)}. 

\bibliography{bib.bib}

\end{document}